\documentclass[12pt]{iopart}
\usepackage[dvips]{graphicx}

\begin{document}

\title[]{The influence of the Kerr effect in Mott insulator to superfluid transition from the point of view of the
Jaynes-Cummings-Hubbard model}

\author{C B Gomes, F A G Almeida and A M C Souza}

\address{Departamento de Fisica, Universidade Federal de Sergipe, 49100-000 Sao Cristovao-SE, Brazil}

\date{\today}

\ead{cbgomes@ufs.br}

\begin{abstract}
We have studied the Jaynes-Cummings-Hubbard model for a chain with the Kerr effect (nonlinear optical effect) through fermionic approximation. We have observed that the Kerr effect does not cause major changes in the energy spectrum. However, the phase transition properties from Mott insulator to superfluid undergoes significant changes due to the Kerr effect.
\end{abstract}


\maketitle

\section{Introduction} \label{sec:int}

In recent years, experimental success in the engineering of high-quality micro-cavities with interactions between photons and
atoms has opened up the possibility of using light-matter systems as quantum simulators for ultracold atoms in optical lattices
\cite{WLI, BDZ, JSGM}. The simplest system of radiation-matter coupling is the interaction of a two-level atom with a single
quantized mode of an optical cavity. An effective model for such interactions was proposed by Jaynes and Cummings (Jaynes-Cummings
model, JCM) 50 years ago \cite{JC,BXYX, BA}.

Recently, a generalization of the JCM has been proposed in order to approach photon-hopping between cavities in optical lattices
\cite{SGB}. Essentially, this generalization, called the Jaynes-Cummings-Hubbard Model (JCHM), describes the competition between
the strong atom-photon coupling and the photon-hopping between cavities. In particular, a quantum phase diagram involving the
transition from the Mott insulator (MI) phase to the superfluid (SF) phase has been investigated using the JCHM \cite{MOTT, DZ}.

The JCM can be exactly solved within the framework of rotating wave approximation \cite{ZD}. However, JCHM does not yet have an
analytical solution requiring approximation methods. Mering \etal \cite{AMPK} proposed an approach which treats spin operators as
fermionic, allowing theoreticians to solve JCHM in momentum space using a Fourier transform over the bosonic and fermionic
operators.

Nonlinear optical effects are usually observed in an optical lattice \cite{YXDJ,KYTT}. Today, these effects have attracted
significant interest due to the possibility of producing entangled states, an achievement particularly important to the field of
quantum information \cite{DMP,ADK}. Using optical lattices, highly applicable results have been obtained using the Kerr effect
\cite{SJA,APA}. The Kerr is an atomic third order effect that occurs in optical cavities \cite{OAA}, which is often overlooked in
cavities of high quality. Generally, it is important to investigate the consequences of its presence. Studies involving JCM in the
presence of the Kerr effect have been performed, including obtaining the statistics of photons emitted from a
cavity driven by an external laser source \cite{PBJ,AKH,SJ}. In the present paper, we investigate the JCHM in Kerr medium to optical chain to large size using the theoretical approach proposed by Mering \etal \cite{AMPK}. We show
that the presence of the Kerr effect in optical lattices influences the transition from MI to SF, increasing critical hopping.

This paper is structured as follows. In \sref{sec:tjchkm} we show the Jaynes-Cummings-Hubbard-Kerr
model describing its effective Hamiltonian. We introduce the
fermionic approximation in \sref{sec:fapp}. In \sref{sec:resu} we expose the results. Finally, our conclusions are
in \sref{sec:conc}. 

\section{The Jaynes-Cummings-Hubbard-Kerr Model} \label{sec:tjchkm}

The Hamiltonian of the Jaynes-Cummings-Hubbard-Kerr Model (JCHKM) to a chain of $L$ atoms is given by ($\hbar=1$)
\begin{eqnarray}
\hat{H} & = &\sum_{j}\omega_{j}\hat{a}_{j}^{\dagger}\hat{a}_{j}+ \epsilon \sum_{j} \hat{\sigma}_{j}^{
+}\hat{\sigma}_{j}^{-}+g\sum_{j}(\hat{a}_{j}^{\dagger}\hat{\sigma}_{j}^{-}+\hat{a}_{j}\hat{\sigma}_{
j}^{+}) \nonumber \\
{} & {}
& +\sum_{d}t_{d}\sum_{j}(\hat{a}_{j}^{\dagger}\hat{a}_{j+d}+\hat{a}_{j+d}^{\dagger}\hat{a}_{j})+
\gamma \sum_{j} \hat{a}_{j}^{\dagger 2} \hat{a}_{j}^{2}.\label{E1} 
\end{eqnarray}
Here $\hat{\sigma}^{\pm}=\hat{\sigma}_{x}\pm i\hat{\sigma}_{y}$ and $\hat{\sigma}_{x,y,z}$ are Pauli
matrices, and $\hat{a}_{j}$ ($\hat{a}^{\dagger}_{j}$) is the annihilation (creation) operator
of the light mode in the $j$th cavity with frequency $\omega_{j}$. 
The frequency of the transition energy of the atoms is denoted by $\epsilon$.
The light-atom coupling is represented by $g$, and $t_{d}$ is the hopping integral
between $d$th-neighboring cavities. The first four terms in
Hamiltonian \eref{E1} represent the JCHM, and the last one is the Kerr term where $\gamma$ is the
constant related to the nonlinear response of the Kerr medium \cite{MWR,BG,XXL,RHX}. 

For a large optical chain we can write $\omega_j = - \omega \equiv 2t\zeta(3)$ and $t_{d}=\frac{t}{d^3}$, where
$t \equiv \frac{\omega_{z}^{2}}{2\omega_{x}\tilde{u}^{3}}$, $\omega_z$ and $\omega_x$ are the longitudinal and transversal
frequencies of light, respectively, and $\tilde{u}$ is the mean equilibrium distance between cavities \cite{DFVJ}.

When the hopping $t_{d}=0$ and Kerr $\gamma=0$ terms vanish, the Hamiltonian \eref{E1} is decoupled into $L$
independent JCM Hamiltonians, which has well-known eigenstates \cite{CNA}. If $\gamma \neq 0$, the system remains decoupled
with the same eigenstates. However, when
$t_d \neq 0$, the cavities become coupled, increasing the complexity of the solution due to the fact that we cannot write the
eigenstates of the whole system as a direct product of single-cavity eigenstates. In this situation, an appropriate approach is
the fermionic treatment \cite{AMPK} followed by a mean field approximation that disregards the momentum transfer between photons
in the Kerr term.

\section {Fermionic Approximation} \label{sec:fapp}

Fermionic treatment consists in replacing the spin operators with fermionic ones. This approach produces exact
results for the JCM. However, as observed in the previous section, the eigenstates of JCHM are not a direct product of
single-cavity JCM eigenstates. In this situation, the present treatment is an approximation that has presented results very close
to other methods such as the mean field approach \cite{AMPK}. We can write Hamiltonian \eref{E1} as

\begin{eqnarray}
\hat{H} & = &\omega\sum_{j}\hat{a}_{j}^{\dagger}\hat{a}_{j}+\epsilon\sum_{j}\hat{c}_{j}^{\dagger}\hat{c}_{j}+g\sum_{j}(\hat{a}_{j}^{\dagger}\hat{c}_{j}+\hat{a}_{j}\hat{c}_{
j}^{\dagger}) \nonumber \\
{} & {}
& +\sum_{d}t_{d}\sum_{j}(\hat{a}_{j}^{\dagger}\hat{a}_{j+d}+\hat{a}_{j+d}^{\dagger}\hat{a}_{j})+
\gamma \sum_{j} \hat{a}_{j}^{\dagger 2} \hat{a}_{j}^{2} \nonumber
\end{eqnarray}
where $\hat{\sigma}^+$ ($\hat{\sigma}^-$) are replaced by fermionic operators $\hat{c}^\dagger$ ($\hat{c}$). Assuming that the
number of cavities is denoted by $L$, and then performing a Fourier transform,
\[
\hat{a}_{j}=\frac{1}{\sqrt{L}}\sum_{k}e^{-2\pi i \frac{kj}{L}}\hat{a}_{k},
\hspace{1cm} \hat{c}_{j}=\frac{1}{\sqrt{L}}\sum_{k}e^{-2\pi i \frac{kj}{L}}\hat{c}_{k}, 
\]
we find
\begin{eqnarray}
\hat{H}&=&\sum_{k}\omega_{k}\hat{a}_{k}^{\dagger}\hat{a}_{k}+g\sum_{k}(\hat{a}_{k}^{\dagger}
\hat{c}_{k}+\hat{a}_{k}\hat{c}_{k}^{\dagger}) \nonumber \\
{} & {} & +\epsilon\sum_{k}\hat{c}_{k}^{\dagger}\hat{c}_{k} +\frac{\gamma}{L}\sum_{k,k',q}
\hat{a}_{k+q}^{\dagger}\hat{a}_{k'-q}^{\dagger} \hat{a}_{k}\hat{a}_{k'}
\label{E3} 
\end{eqnarray}
where $\omega_{k} \equiv 2t\left[\sum_{d=1}^{\infty} \frac{cos(2\pi \frac{kd}{L})}{d^3} -\zeta(3)
\right]$. 

Due to the Kerr term, the Hamiltonian \eref{E3} cannot be written as a $k$ summation. Subsequently, we employ a mean field
approximation to decouple this term, assuming that $q = 0$ or that, in other words, there
is no momentum exchange. Under these circumstances, we will see that the solution of the Hamiltonian depends on self-consistent
equations similar to the standard mean field approach. The last term of the Hamiltonian \eref{E3} is written as
\[
 \frac{\gamma}{L} \sum_{k k'}  \hat{a}_k^\dagger \hat{a}_{k'}^{\dagger} \hat{a}_k \hat{a}_{k'} = 
 \frac{\gamma}{L} \sum_{k k'}  \hat{a}_k^\dagger ( \hat{a}_k \hat{a}_{k'}^{\dagger} -\delta_{k k'}  ) \hat{a}_{k'} = 
 \gamma \left( \hat{\bar{n}} - \frac{1}{L} \right) \sum_k  \hat{a}_k^\dagger \hat{a}_{k},
\]
where $\hat{\bar{n}} \equiv \frac{1}{L} \sum_k  \hat{a}_k^\dagger \hat{a}_{k}$ is the photon-density operator of the system.
Therefore, in the large chain limit ($L \gg 1$) we can write the Hamiltonian as 
\[
 \hat{H} = \sum_k \left[ \hat{\bar \omega}_k  \hat{a}_k^\dagger \hat{a}_k + \epsilon \hat{c}_k^\dagger \hat{c}_k + g (\hat{a}_k^\dagger
\hat{c}_{k}+\hat{a}_{k}\hat{c}_{k}^{\dagger})  \right].
\]
in which $\hat{\bar \omega}_k \equiv \omega_k +\gamma \hat{\bar{n}}$. Because we are interested in obtaining transition points on
the phase diagram,
we employ the fact that, in the MI phase, the mean number of photons in the ground state, $ n_0$, is the same for all moments, and
consequently 
$\langle \hat{\bar{n}} \rangle_0 = \frac{1}{L} \sum_k  \langle \hat{a}_k^\dagger \hat{a}_{k} \rangle_0 = n_0$ \cite{AMPK}. Now, 
by the follow replacement method, $\hat{\bar{n}} \rightarrow n_0$, and we obtain $\hat{H} = \sum_k \hat{H}_k$, in which
\begin{eqnarray}
 \hat{H}_k & = & \bar{\omega}_k  \hat{a}_k^\dagger \hat{a}_k + \epsilon \hat{c}_k^\dagger \hat{c}_k + g (\hat{a}_k^\dagger
\hat{c}_{k}+\hat{a}_{k}\hat{c}_{k}^{\dagger}), \label{hk} \\
\bar{\omega}_k & = & \omega_k + \gamma n_0. \label{wk}
\end{eqnarray}
In this way, the system is decoupled and its eigenstates become direct products of the individual $k$ eigenstates, allowing us to
study our system through Hamiltonian \eref{hk}. We assume the notation $| l,m \rangle$ for the state with $l$ fermions and $m$ photons with $k$ momentum. Note that the total number
 of excitations, $\hat{N}_k = \hat{a}_k^\dagger \hat{a}_k + \hat{c}_k^\dagger \hat{c}_k$, commutes with $\hat{H}_k$. 
Therefore, we resorting to the basis $\{ |0,n \rangle, |1,n-1 \rangle \}$ because the total number of
excitations is constant, and the ground state of the Hamiltonian \eref{hk} is given by
\[
 | \psi_0 \rangle = \alpha | 0,n \rangle + \beta | 1,n-1 \rangle,
\]
in which
\[
 \alpha \equiv \frac{g \sqrt{n}}{\sqrt{2 \chi_k^n} \sqrt{ \chi_k^n +
\frac{\bar{\omega}_k-\epsilon}{2}}}, \hspace{1cm}
 \beta \equiv \frac{\sqrt{ \chi_k^n + \frac{\bar{\omega}_k-\epsilon}{2}}}{\sqrt{2 \chi_k^n}},
\label{beta}
\]
and
\[
 \chi_k^n \equiv \sqrt{g^2 n + \left( \frac{\bar{\omega}_k-\epsilon}{2} \right)^2}.
\]

Afterwards, it is relatively easy to find that the ground state energy of Hamiltonian 
It is easy to find that the ground state energy of the Hamiltonian \eref{hk} is 

\begin{equation}
E_k^n=\bar{\omega}_{k}n-\frac{(\bar{\omega}_{k} - \epsilon)}{2} - \chi_k^n .
\label{E4}
\end{equation}
Now, we can compute the mean photon number in the ground state as follows
\begin{equation}
 n_0 =  \langle \psi_0 |\hat{a}_k^\dagger \hat{a}_k | \psi_0 \rangle = n |\alpha|^2 + (n-1) |\beta|^2 = n - |\beta|^2 \label{n0}.
\end{equation}
The self-consistent mean field equation is closed by inserting \eref{n0} in \eref{wk}.

When hopping, the term is non-null in the JCHM, and there is a transition between the MI and SF states \cite{AMPK}. The equation
\eref{E4} allows us to investigate the influence of the Kerr effect in this transition. 

\section{MI-SF transition} \label{sec:resu}

\begin{figure}
\centering
\includegraphics[scale=0.49]{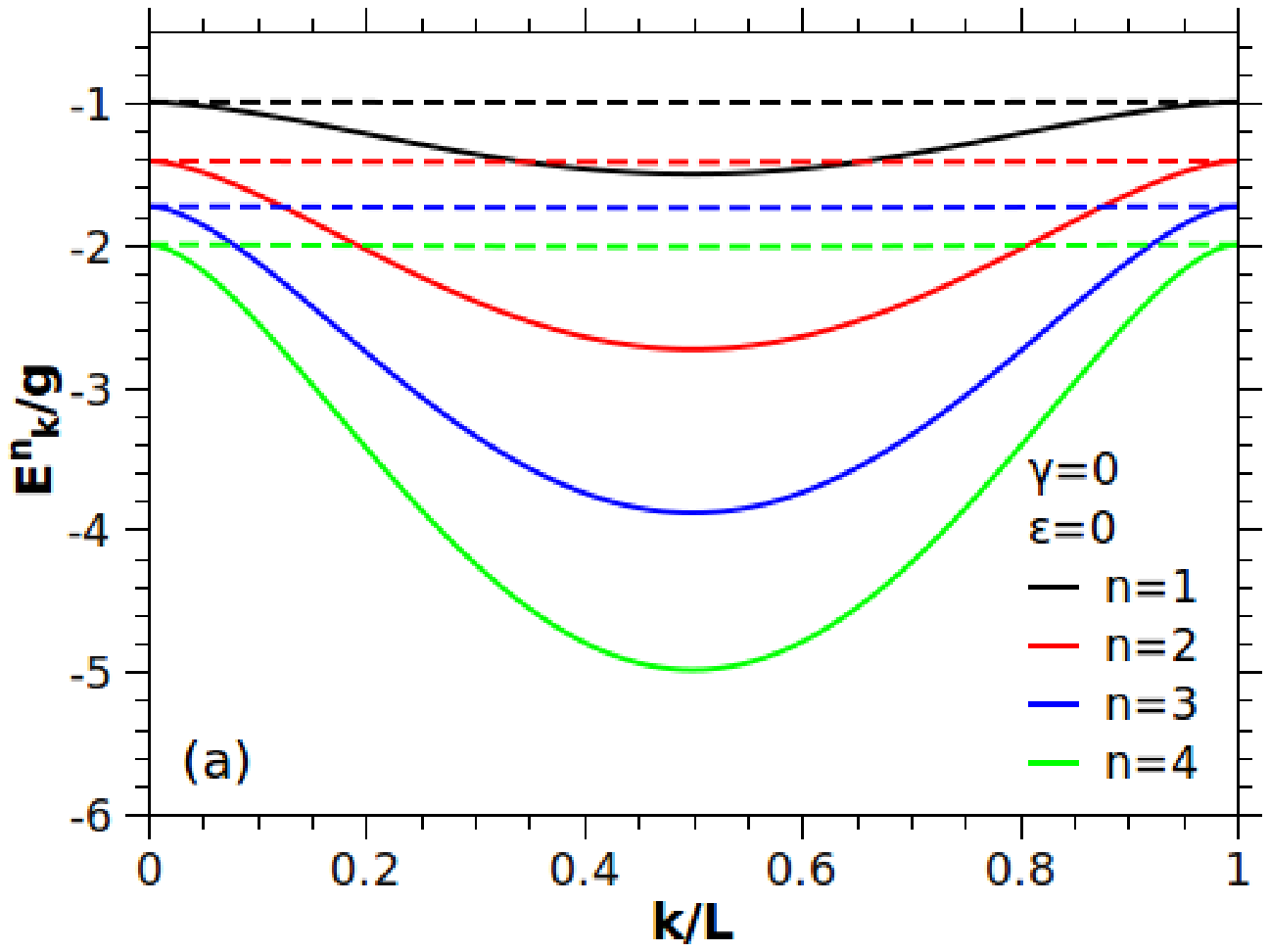} \includegraphics[scale=0.49]{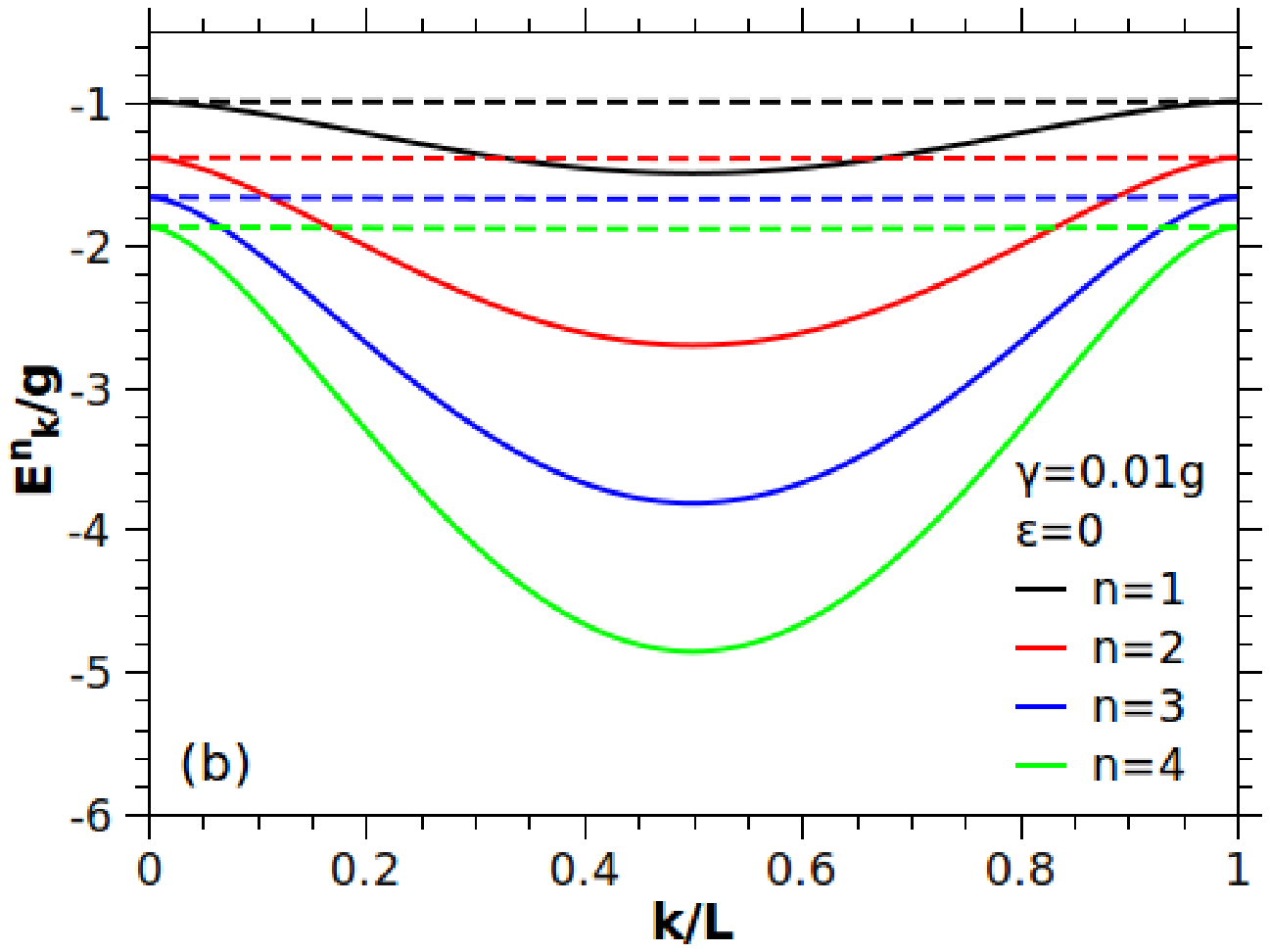}
\includegraphics[scale=0.49]{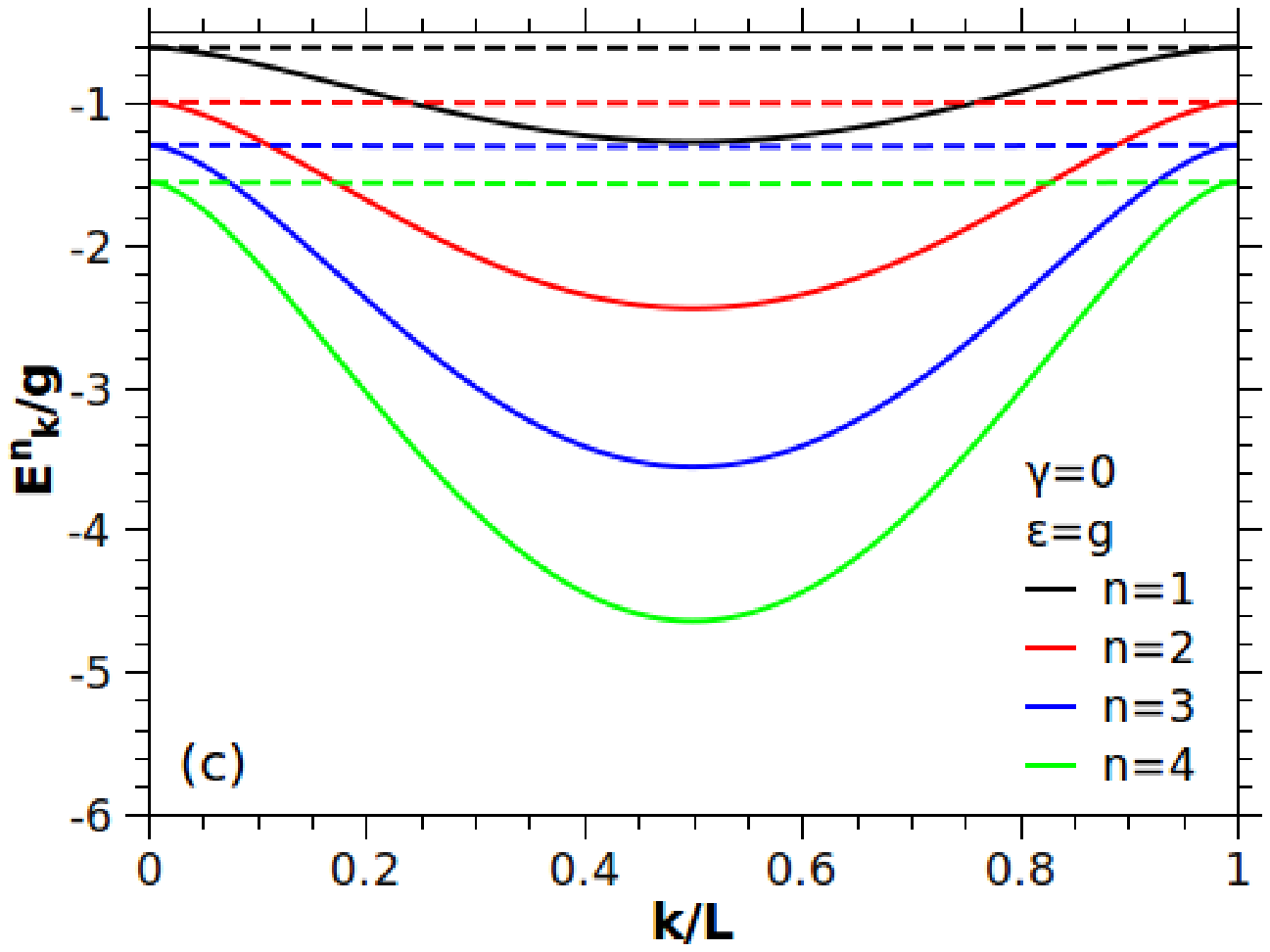} \includegraphics[scale=0.49]{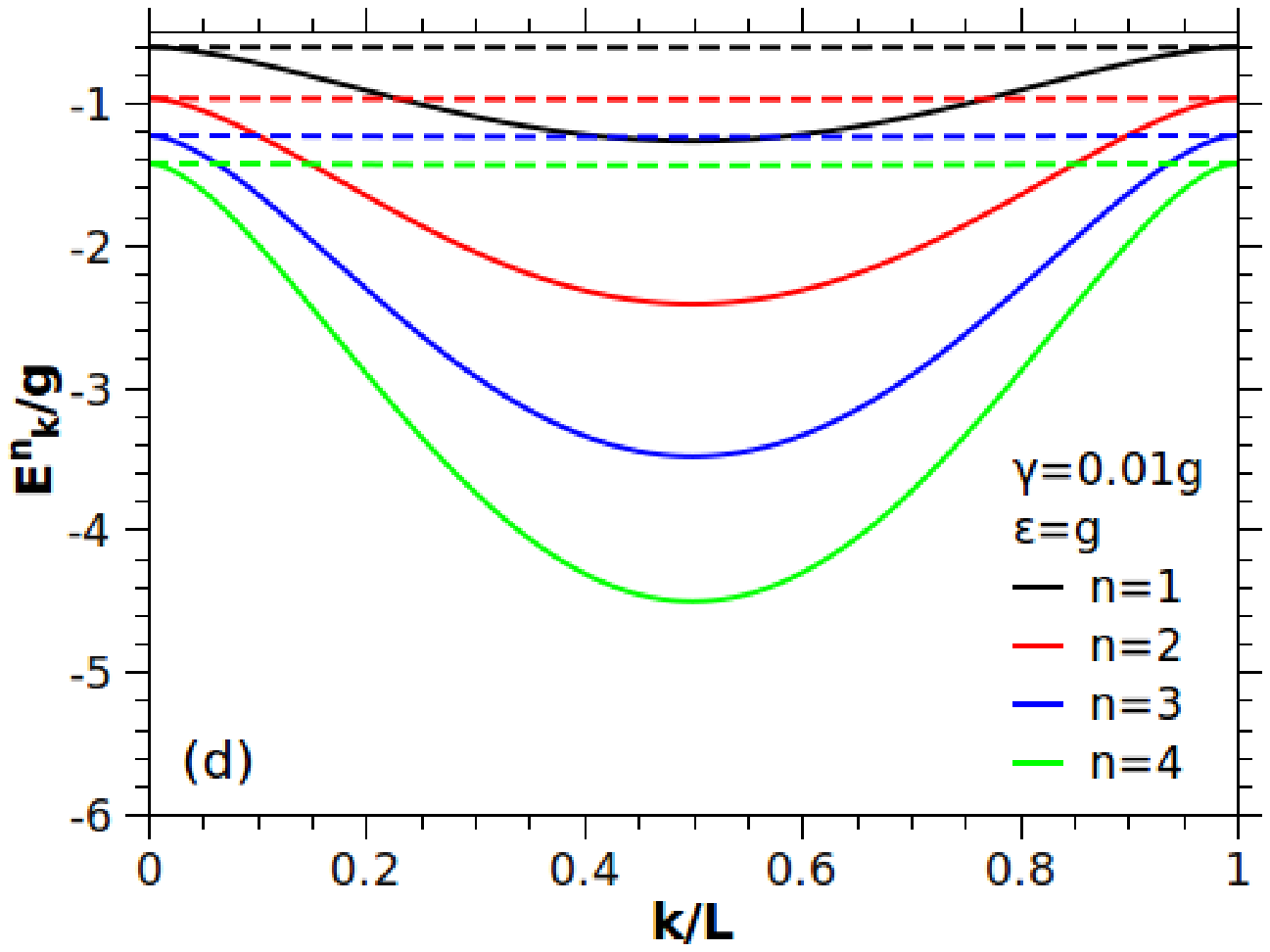}
\caption{
Energies for typical values of $\epsilon$ and $\gamma$ are given by \eref{E4}. Solid
lines refer to $t/g=0.2$ while dashed ones are related to $t/g=0.001$. Note that when
$t/g\rightarrow0$, the minimum disappears.}
\label{fig1}
\end{figure}

\Fref{fig1} shows how energy depends on $k$. Observe that for $t \rightarrow 0$ the energy has no dependence on $k$. Consequently,
the photons are uniformly distributed in momentum space characterizing a MI phase, which has a gap between the chemical potentials
of the particles and the hole \cite{AMPK}. However, the gap is reduced as $t$ increases until it becomes null at a specific
critical point, $t_c$, appearing in the transition from the MI to SF phases. The chemical potential of the
particle $\mu_{n}^{+}=E_{k'}^{n+1}-E_{k'}^{n}$, while the hole's potential $\mu_{n}^{-}=E_{k}^{n}-E_{k}^{n-1}$,
where $k'$ ($k$) is the minimum (maximum) of energy \eref{E4}. In \fref{fig1}, $k'=L/2$
and
$k=0$ or $k=L$ for any parameters $n$ and $\gamma$. Therefore, the chemical potentials are calculated
through the following expressions
\begin{eqnarray}
  \mu_{n}^{+}=E_{L/2}^{n+1}-E_{L/2}^{n}, \hspace{1cm}  \mu_{n}^{-}=E_{0}^{n}-E_{0}^{n-1} .
\label{E5}
\end{eqnarray}

\begin{figure}
\centering
\includegraphics[scale=0.49]{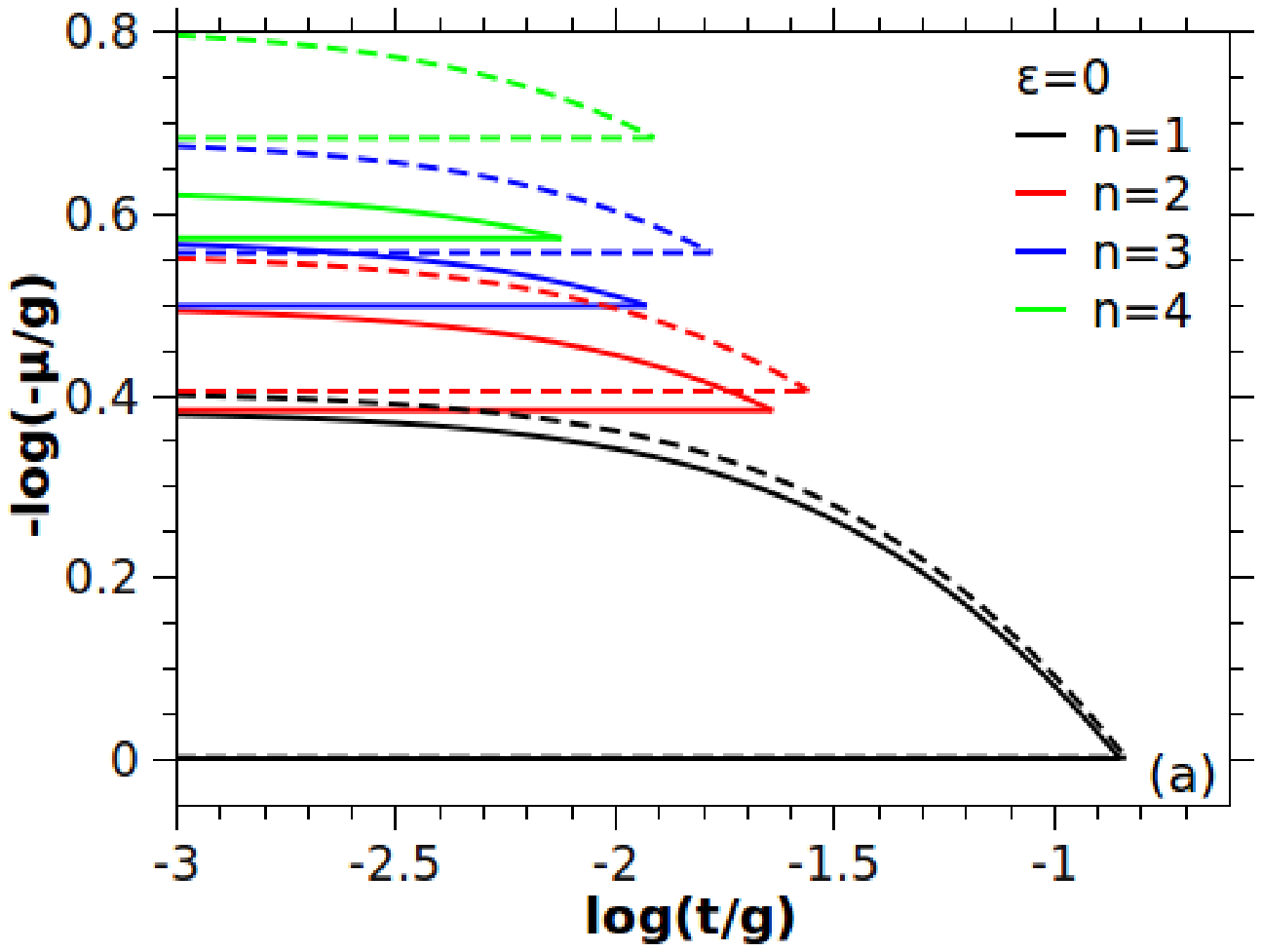}
\includegraphics[scale=0.49]{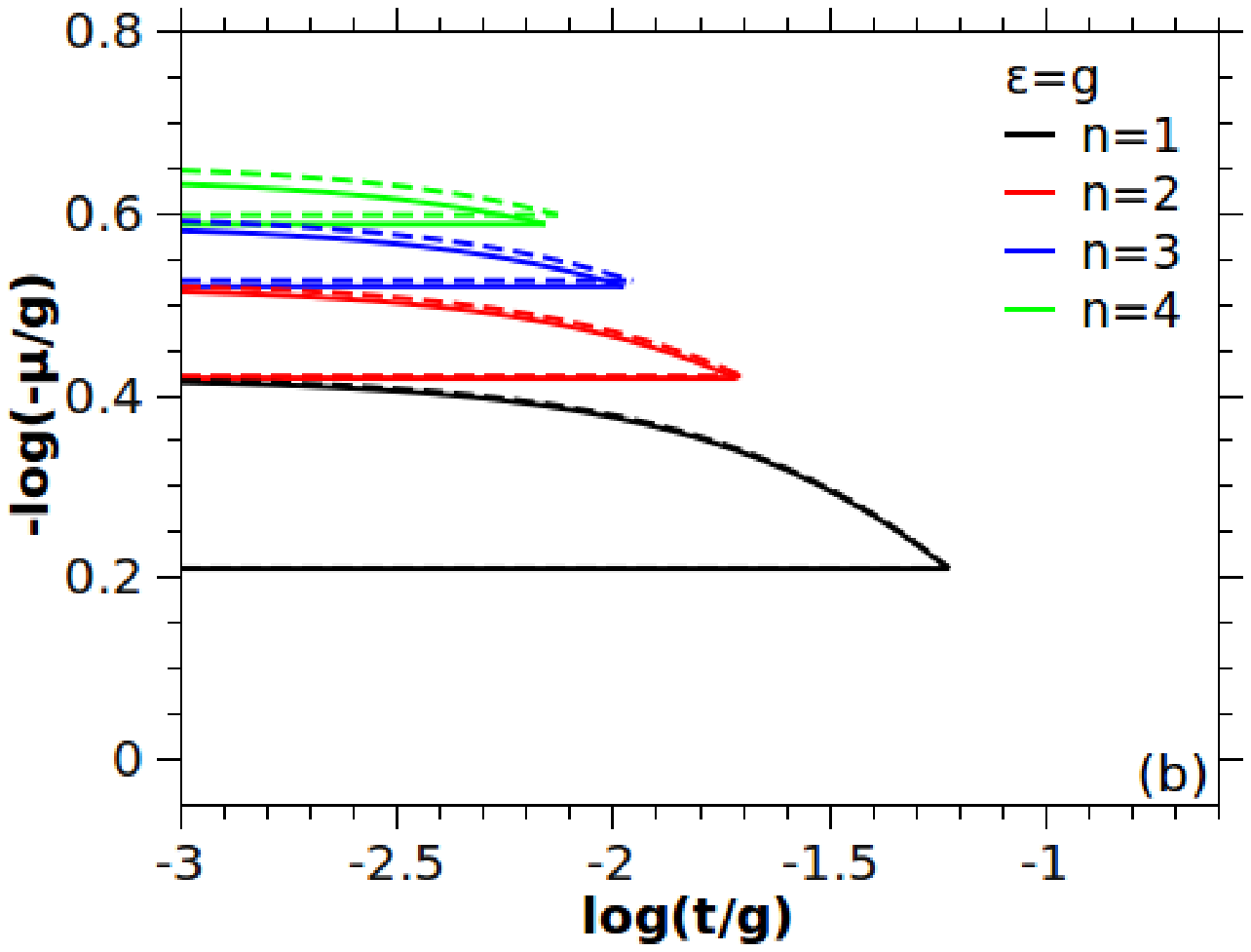}
\caption{Mott lobes for (a) $\epsilon=0$ and (b) $\epsilon=g$. ashed lines refer to $\gamma=0$
while solid ones are related to $\gamma/g=0.01$. } \label{fig2}
\end{figure}

The well-known feature of the MI-SF transition is the Mott lobe which is exhibited in \Fref{fig2} 
for some values of $n$, $\epsilon$, and $\gamma$.
The plus (minus) sign in \eref{E5} corresponds to the upper (lower) boundary of the Mott lobe. The Kerr effect produces a shift of
Mott lobes, which is emphasized as $n$ increases. This result is expected because the Kerr term in Hamiltonian \eref{E1} has a
$n^2$ dependence. The lobs are also shifted through changes in $\epsilon$ in agreement with the results of non-Kerr
models \cite{AMPK}. When
$\mu_{n}^{+}=\mu_{n}^{-}$ the lobe is closed at the critical point, $t_c$, which depends on $n$,
$\gamma$ and $\epsilon$. 
\begin{figure}
\centering
\includegraphics[scale=0.50]{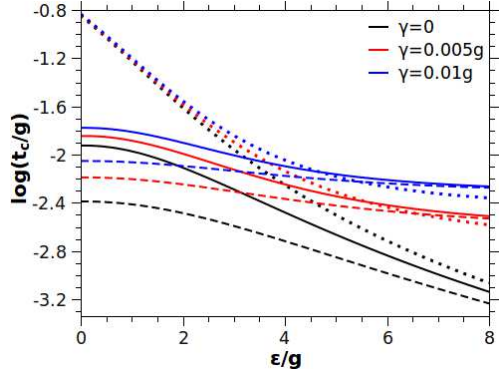}
\caption{Critical hopping in terms of $\epsilon$ to some $\gamma$ values, and $n =$ 1
(dotted lines), 3 (solid lines), and 6 (dashed lines).}\label{fig3}
\end{figure}

The dependence of $t_c$ on $\epsilon$ for some values of $n$ and $\gamma$ is exhibited in
\fref{fig3}. Note that $t_c$ decreases as $\epsilon$ increases meaning that critical hopping is smaller as atomic-level spacing
increases. Furthermore, the increase of $n$ reduces $t_c$ in accordance with non-Kerr model results \cite{AMPK}. Additionally,
$t_c$ increases with $\gamma$, and this behavior is amplified as large as $n$, which is expected due to the $n^2$ dependence of
the Kerr term in Hamiltonian \eref{E1}.

\section{Conclusion} \label{sec:conc}

We have studied properties of the MI-SF transition on a
long-range-hopping JCHK chain through fermionic
approximation followed by a mean field approach. Despite the fact that the Kerr effect be very small ($\gamma \ll g$,
see  \sref{sec:int}) and it
causes negligible changes in the energy spectrum (see \fref{fig1}), it significantly changes the
properties of the transition (see \fref{fig2} and \fref{fig3}). A similar feature had been exhibited in reference \cite{HAPS}
where it can be observed that changes in the photon hopping range influence slightly the energy spectrum while modifying
significantly MI-SF transition properties. Therefore, we wish to emphasize the importance of taking into account properties that
induce small effects in the energy spectrum when the research focus is MI-SF transition properties.

\ack
This work was supported by CAPES, FAPITEC/SE and CNPq (Brazilian Agencies).


\section*{References}

\begin{thebibliography}{99}

\bibitem{WLI} Li W, Hamadeh L and Lesanovsky I 2012 Phys. Rev. A 85 053615.
\bibitem{BDZ} Bloch I, Dalibard J, and Zwerger W 2008 Rev. Mod. Phys. 80 885.
\bibitem{JSGM} Jordens R \etal 2008 Nature 455 204-207.
\bibitem{JC} Jaynes E T Cummings F W 1963 Proc. IEEE 51: 89-109.
\bibitem{BXYX} Bang-Fu D \etal 2011 Commun. Theor. Phys. 55 662.
\bibitem{BA} Bougouffa S and Al-Awfi S 2009 Phys. Scr. T135 014011.
\bibitem{SGB} Schmidt S and Blatter G 2009 Phys. Rev. Lett 103 086403.
\bibitem{MOTT} Greiner M \etal 2002 Nature (London) 415 39.
\bibitem{DZ} Dzyaloshinskii I 1989 Phys. Scr. Vol. T27 89-95.
\bibitem{ZD} Zueco D \etal 2009 Phys. Rev. A 80 033846.
\bibitem{AMPK} Mering A \etal 2009 Phys. Rev. A 80 053821.
\bibitem{YXDJ} He Y \etal 2012 Phys. Rev. A 85  013831.
\bibitem{KYTT} Eguchi K \etal 2012 Phys. Rev. B 85 174415.
\bibitem{DMP} Vitali D, Fortunato M and Tombesi P 2000 Phys. Rev. Lett. 85 445.
\bibitem{ADK} Angelakis D G, Dai L, and Kwek L C 2010 Europhys. Lett. 91 10003.
\bibitem{SJA} Schoenes J 1993 Phys. Scr. 1993 289
\bibitem{APA} Adam P \etal 2011 Phys. Scr. 2011 014002
\bibitem{OAA} Obada A F \etal 1998 Eur. Phys. J. D  3 3 289-294.
\bibitem{PBJ} Patargias N, Bartzis V and Jannussist A 1995 Phys. Scr. 52 554-557.
\bibitem{AKH} Bu S \etal 2008 Phys. Scr. 78 065008.
\bibitem{SJ} Cordero S and Recamier J 2011 J. Phys. B: At. Mol. Opt. Phys. 44 135502.
\bibitem{MWR} Werner M J and Risken H 1991 Phys. Rev. A 44 Number 7.
\bibitem{BG} Bandyopadhyay A and GangopadhyaY G 1996 J. Mod. Opt. 43 487.
\bibitem{XXL} Xie R H, Xu G O and Liu D H 1995 Aust. J. Phys. 48 907.
\bibitem{RHX} Xie R H 1996 Can. J. Phys. 74 305.
\bibitem{CNA} Nietner C and Pelster A 2012 Phys. Rev. A 85 043831.
\bibitem{DFVJ} James D F V 1998 Appl. Phys. B: Lasers Opt. 66 181.
\bibitem{HAPS} Hohenadler M \etal 2012 Phys. Rev. A 85 013810.

\end{thebibliography}
\end{document}